\documentclass{ws-ijmpe}

\def\ga{\mathrel{\mathchoice {\vcenter{\offinterlineskip\halign{\hfil
$\displaystyle##$\hfil\cr>\cr\sim\cr}}}
{\vcenter{\offinterlineskip\halign{\hfil$\textstyle##$\hfil\cr>\cr\sim\cr}}}
{\vcenter{\offinterlineskip\halign{\hfil$\scriptstyle##$\hfil\cr>\cr\sim\cr}}}
{\vcenter{\offinterlineskip\halign{\hfil$\scriptscriptstyle##$\hfil\cr>\cr
\sim\cr}}}}}

\begin{document}

\markboth{K.-H. Kampert}{Cosmic Rays at the highest energies - 
First data from the Pierre Auger Observatory}

\catchline{}{}{}{}{}

\title{COSMIC RAYS AT THE HIGHEST ENERGIES\\
-- FIRST DATA FROM THE PIERRE AUGER OBSERVATORY --}

\author{\footnotesize KARL-HEINZ KAMPERT}

\address{University of Wuppertal\\
Department of Physics, Gau{\ss}str.\ 20\\
D-42117 Wuppertal, Germany\\
{\tt kampert@uni-wuppertal.de}}

\author{FOR THE PIERRE AUGER COLLABORATION}

 
\maketitle

\begin{history}
\received{(received date)}
\revised{(revised date)}
\end{history}

\begin{abstract} 
The southern Pierre Auger Observatory, presently under
construction in Malarg\"ue, Argentina, is nearing completion.
The instrument is designed to measure extensive air-showers with
energies ranging from $10^{18}$-$10^{20}$~eV and beyond.  It
combines two complementary observation techniques; the detection
of particles at ground and the coincident observation of
associated fluorescence light generated in the atmosphere above
the ground.  This is being realized by employing an array of 1600
water Cherenkov detectors, distributed over an area of
3000~km$^{2}$, and operating 24 wide-angle Schmidt telescopes,
positioned at four sites at the border of the ground array.  The
Observatory will reach its full size only in 2007 but data are
routinely recorded already and have started to provide relevant
science results.  This talk will focus on the detector
characterizations and presents first results on the arrival
direction of extremely-high energy cosmic rays, their
energy spectrum, and on the upper limit of the photon fraction.
\vspace{1pc}
\end{abstract}

\section{Introduction}
Over the past decade, interest in the nature and origin of
extremely high energy cosmic rays (EHECR) has grown enormously.
Of particular interest are cosmic rays (CR) with energies $\ga
10^{20}$~eV. There is a twofold motivation for studying this
energy regime, one coming from particle physics because CRs give
access to elementary interactions at energies much higher than
man-made accelerators can reach, and another coming from
astrophysics, because we do not know what kind of particles they
are and where and how they acquire such enormous energies.  An
excellent review, published by Michael Hillas 20 years ago,
presented the basic requirements for particle acceleration to
energies $\ge 10^{19}$~eV by astrophysical
objects.\cite{Hillas84} The requirements are not easily met,
which has stimulated the production of a large number of creative
papers.

The problem is aggravated even more by the fact that at these
energies protons and nuclei should interact with the Cosmic
Microwave Background (CMB).  Above a threshold energy of $E_{\rm
GZK}\simeq 5 \times 10^{19}$~eV protons lose their energy over
relatively short cosmological distances via photo-pion production
$p+\gamma_{\rm CMB} \to \pi^{0}+p$ or $\pi^{+}+n$.
Accidentally, nuclei (He, \ldots Fe) lose their energy at similar
threshold energies and on even shorter length scales.  This is
because of photodissociation (e.g. ${\rm Fe} + \gamma_{\rm CMB}
\to X + n$) taking place mostly via giant nuclear resonances.
Finally, photons interact even more rapidly in the CMB by
producing $e^{+}e^{-}$-pairs.  Thus, particles that have traveled
over distances of 50 or 100 Mpc are unlikely to retain an energy
of $\sim 10^{20}$~eV or more when they reach us.  This was
already recognized in the 1960's shortly after the discovery of
the CMB and is called the Greisen-Zatsepin-Kuzmin (GZK)
cutoff.\cite{GZK} Thus, not only do we not know how particles
could obtain such extreme energies even in the most powerful
astrophysical accelerators, these accelerators have to be located
nearby on cosmological scales!

To solve this most pressing puzzle of high energy astroparticle
physics, one either needs to invent nearby exotic EHECR sources
or find ways of evading the GZK effect.  Top-Down models with
decaying topological defects or decaying superheavy relic
particles are typical representatives of the former group, as
EHECRs would be produced nearby.  Typical representatives of the
latter kind are violation of the Lorentz invariance, propagation
of heavy supersymmetric particles, or the $Z$-burst model.  A
comprehensive review, with emphasis placed on top-down models, is
given by Ref.~\cite{Bhattacharjee00}.  Generally, the top-down
models predict a dominance of photons and neutrinos over protons
or nuclei, so that measurements of the chemical composition
become important also at the highest energies.  Furthermore, the
$Z$-burst model cannot avoid producing a strong background of GeV
energy photons leading to severe constrains due to the measured
EGRET fluxes.\cite{Sigl04} Such complications have recently given
more emphasis again to astrophysical sources.

While the large magnetic rigidity of $\sim 10^{20}$ eV protons
gives rise to the problems of particle acceleration in
astrophysical sources, it opens at the same time a new window for
astronomy with CRs.  Since such particles cannot deviate
much in the magnetic fields of the Galaxy and extragalactic
space, they should point to their sources within a few degrees
deviation only.  For example, using nominal guesses of 1~nG for
the magnetic field strength of extragalactic space and 1 Mpc for
the coherence length, deviations for protons on the order of
$2.5^\circ$ are expected after travelling 50 Mpc.\cite{Cronin92}

Two types of experiments based on very different techniques have
undoubtedly detected particles well exceeding the GZK
cut-off.\cite{NaganoWatson00,Takeda03,Abbasi04a} Unfortunately,
despite 40 years of data taking the number of events is still
small.  Also, the largest experiments so far disagree at an
approx.\ $2 \sigma$ level on the flux and on arrival direction
correlations.  The HiRes collaboration, employing the
fluorescence technique, reported a suppression of the flux above
the GZK-threshold, with no evidence for clustering in the arrival
directions.\cite{Abbasi04a,Abbasi04b} On the other hand, ground
arrays have detected no GZK-cutoff.\cite{NaganoWatson00,Takeda03}
Furthermore, the the AGASA collaboration published results about
seeing a clustering of the highest energy events \cite{Takeda03}
which, however, is not free of dispute.\cite{Finley04} Clearly,
the situation is very puzzling, and a larger sample of high
quality data is needed for the field to advance.

\section{The Pierre Auger Observatory}

Already years before the present controversy between different
experiments started, it was clear that not only a much larger
experiment was needed to improve the statistics of EHECRs on
reasonable time scales but also that two or more complementary
experimental approaches had to be combined on a shower-by-shower
basis within one experiment.  Such redundancy allows
cross-correlations between experimental techniques, thereby
controlling the systematic uncertainties.  Furthermore, one
expects to improve the resolution of the energy, mass, and
direction of reconstructed primary particles.  In the Pierre
Auger Observatory, this so-called `hybrid' aspect is realized by
combining a ground array of water Cherenkov detectors with a set
of fluorescence telescopes.  Another important objective was to
obtain a uniform exposure over the full sky.  This will be
achieved by constructing two instruments, each located at
mid-latitudes in the southern and northern hemispheres.  Each
site is conceived to cover an area of 3000~km$^{2}$ in order to
collect about one event per week and site above $10^{20}$~eV, 
depending on the extrapolation of the flux above the GZK 
threshold.

The ground array will comprise 1600 cylindrical water Cherenkov
tanks of 10~m$^{2}$ surface area and 1.2 m height working
autonomously by solar power and communicating the fully digitized
data by radio links.  The tanks are arranged on a hexagonal grid
with a spacing of 1.5 km yielding full efficiency for extensive
air shower (EAS) detection above $\sim 5\cdot10^{18}$~eV.
Presently, about 1000 tanks are in operation and taking data.

\begin{figure}[t]
\centerline{\epsfxsize=9cm\epsfbox{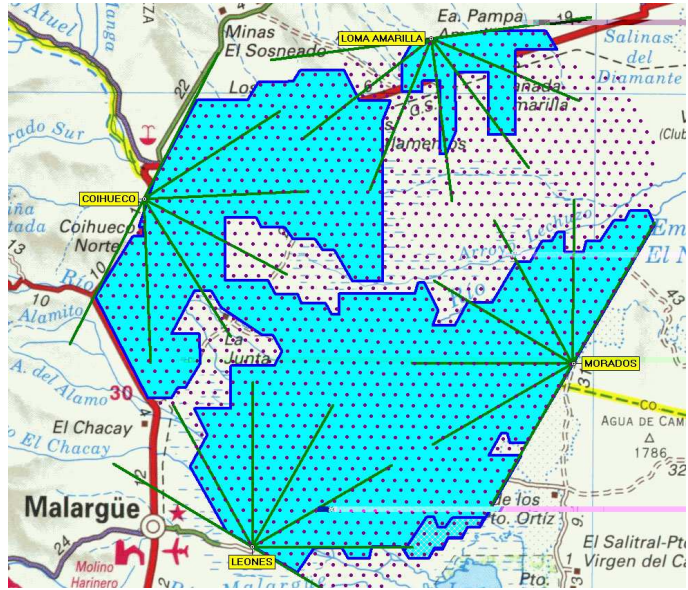}}
\caption[xx]{Layout of the southern site with the locations of
the surface detector tanks indicated.  Also shown are the
locations of the flourescence-eyes with the f.o.v.\ of their
telescopes.  The blue region indicates the part of the ground
array currently in operation (March 2006).  Furthermore, all
telescopes at the Los Leones, Coihueco, and Loma Amarilla site
are in operation.
\label{fig:Site}}
\end{figure}

Charged particles propagating through the atmosphere excite
nitrogen molecules causing the emission of (mostly) ultraviolet
light.  The fluorescence yield is very low, approx.\ four photons
per meter of electron track (see e.g.\,\cite{Kakimoto96}), but
can be measured with large area imaging telescopes during clear
new- to half-moon nights (duty cycle of $\approx$ 10-15\,\%).
The fluorescence detector of the southern site will comprise 24
telescopes arranged into four `eyes' located at the perimeter of the
ground array.  Each eye houses six Schmidt telescopes with a
$30^{\circ} \times 30^{\circ}$ field of view (f.o.v.).  Thus, the
6 telescopes of an eye provide a $180^{\circ}$ view towards the
array center and they look upwards from $1^\circ$ to $31^\circ$
above the horizon.  Presently, 18 telescopes are in operation 
and taking data.

The layout of the southern site and its current status is
depicted in figure\,\ref{fig:Site}.  It shows the locations of
telescopes and water tanks already in operation.  Further details
about the experiment and its performance can be found in Refs.\
\cite{Auger04,Kampert06}.  Nearing completion of the Southern
Site, the collaboration has selected southeast Colorado to site
the northern detector and started to perform related R\&D work.

\section{Anisotropies near the direction of the Galactic Center}

The Galactic Center (GC) region constitutes an attractive target
for CR anisotropy studies at EeV ($10^{18}$\,eV) energies.  These
may be the highest energies for which the galactic component of
the cosmic rays is still dominant.  Moreover, since the GC
harbors a very massive black hole associated with the radio
source Sagittarius A$^{*}$, as well as the expanding supernova
remnant Sagittarius A East, it contains objects that might be
candidates for powerful CR accelerators.  The location of the
Pierre Auger Observatory in the southern hemisphere makes it
particularly suitable for anisotropy studies in this region since
the GC, passing only $6^{\circ}$ from the zenith at the site,
lies well within the field of view of the experiment.  The number
of CRs of EeV energies accumulated so far at the Pierre Auger
Observatory from this part of the sky greatly exceeds that from
previous experiments, allowing several interesting searches to be
made.

As mentioned above, the Akeno Giant Air Shower Array (AGASA)
experiment reported a 4.5 $\sigma$ excess of CRs with energies in
the range $10^{18}$-$10^{18.4}$~eV in a $20^{\circ}$ radius
region centered at right ascension and declination coordinates
$(\alpha,\delta) \simeq
(280^{\circ},-17^{\circ})$.\cite{Hayashida99,Teshima01} The
number of observed and expected events are $n_{\rm obs}/n_{exp} =
506/413.6 = 1.22 \pm 0.05$, where the error quoted is the one
associated with Poisson background fluctuations.  Note that the
GC itself, for which we will adopt hereafter the Sagittarius
A$^{*}$ J2000.0 coordinates, $(\alpha,\delta) =
(266.3^{\circ},-29.0^{\circ})$, lies outside the AGASA field of
view ($\delta>-24.2^\circ$).  A subsequent reanalysis of SUGAR
data failed to confirm these findings, but reported a 2.9
$\sigma$ excess flux of CRs with energies in the range
$10^{17.9}$--$10^{18.5}$~eV in a region of $5.5^\circ$ radius
centered at $(\alpha,\delta)=(274^\circ,-22^\circ)$, for which
they obtained $n_{obs}/n_{exp}=21.8/11.8=1.85\pm
0.29$.\cite{Bellido01}

In order to verify these findings, the arrival directions
measured by the Pierre Auger Observatory data have been analyzed.
We consider the events from the surface detector array with three
or more tanks triggered in a compact configuration.  The events
have to satisfy quality cuts, requiring that the detector with
the highest signal be surrounded by a hexagon of working
detectors.  This ensures that the event is well reconstructed.
We also restrict the events to zenith angles $\theta<60^\circ$.

The energies are obtained using the inferred signal size at
1000~m from the reconstructed shower core, $S(1000)$, adopting a
conversion that leads to a constant flux in different sky
directions above 3~EeV, where the acceptance is saturated.  This
is the so-called `Constant Intensity Cut' criterion implemented
in \cite{so05}.  A calibration of the energies is performed using
clean fluorescence data, i.e.\ hybrid events that were recorded
when there were contemporaneous aerosol measurements, whose
longitudinal profiles include the shower maximum in a measured
range of at least 350~g~cm$^{-2}$ and in which there is less than
10\% Cherenkov contamination.  The estimated systematic
uncertainty in the reconstructed shower energy with the
fluorescence technique is currently 25\%.\cite{be05} ~In this
energy range 48\% of the events involve just three tanks, 34\%
involve 4 tanks and only 18\% more than 4 tanks.  For three tank
events the 68\% quantile angular resolution is about $2.2^\circ$
and the resolution improves for events with 4 tanks or
more.\cite{angres}

\begin{figure}[ht]
\vspace*{-12mm}\centerline{\epsfxsize=14cm\epsfbox{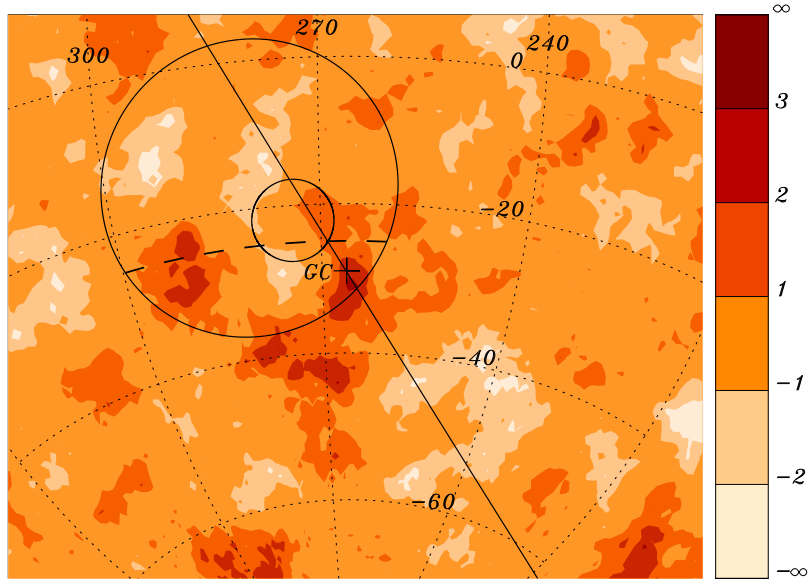}}
\vspace*{-10mm}\caption[xx]{Map of CR overdensity significances
near the GC region on top-hat windows of $5^\circ$
radius.\cite{gcentre-paper} The GC location is indicated by a
cross, lying along the galactic plane (solid line).  Also the
regions where the AGASA experiment found their largest excess
(large circle) as well as the region of the SUGAR excess (small
circle) are indicated.}
\label{gcentre}
\end{figure}

After applying all quality cuts, about 80,000 events remained to
be analyzed in the energy range $10^{17.9}\ {\rm
eV}<E<10^{18.5}$~eV. To study the possible presence of
anisotropies, background expectations for different sky
directions were calculated under the assumption of an isotropic
CR distribution.  This was performed by applying both a
semi-analytic and a shuffling technique.  Both methods were found
to agree within 0.5\,\%, i.e. within their statistical
fluctuations.  Figure~\ref{gcentre} shows the resulting map of
the GC region in terms of the so called Li-Ma significances
\cite{li83} of overdensities in circular windows of $5^{\circ}$
radius and for the aforementioned energy range.  This angular
scale is convenient to visualize the distribution of
overdensities in the windows explored by SUGAR and AGASA. The
galactic plane is represented by a solid line and the location
of the Galactic Center is indicated by a cross.  The region in
which AGASA reported an excess (in a slightly narrower energy
range) is the big circle in the neighborhood of the GC, with the
dashed line indicating the lower boundary of the region observed
by AGASA. The smaller circle indicates the region where an excess
in the SUGAR data was reported.

The size of the overdensities present in this map is consistent
with what would be expected as a result of statistical
fluctuations of an isotropic sky.  Indeed, inspecting the
distribution of these overdensities together with the
expectations from an isotropic flux (average and $2\sigma$ bounds
obtained from Monte Carlo simulations), does not show any
significant departure from isotropy.  For the $20^\circ$ circle
centered at the AGASA location and for $10^{18}\ {\rm
eV}<E<10^{18.4}$~eV, 2116 events are observed while 2159.6 are
expected using the semi-analytic technique, while 2169.7 are
expected using the shuffling technique.  Note that the number of
events is more than four times that collected by AGASA in this
region, in part due to the fact that the GC lies well within the
field of view of Auger, and in part due to the fact that the
total exposure of Auger is already double that achieved by AGASA.
The largest source of systematic uncertainties when comparing the
AGASA and Auger excess maps may be given by the uncertainties of
their energy scales.  To test the effect of this, the Auger
energy scale was shifted by $\log(E)=\pm0.1$ and the anisotropy
analysis repeated.  Independent of such shifts, the excess is
always compatible with zero.

Regarding the localized excess observed in SUGAR data, we find
$n_{obs}/n_{exp}=286/289.7=0.98\pm 0.06$ in the same angular
window and energy range.  Hence, with more than an order of
magnitude larger statistics no significant excess is seen in this
window.  Shifting the energy range to account for possible
offsets again resulted in no significant excess.

To complete the analysis of the GC region, we have also searched
for a point like source located in the position of Sagittarius
A$^*$.  For a Gaussian window corresponding to the angular
resolution of the experiment we get $n_{obs}/n_{exp}= 53.8/45.8$.
This corresponds to a ratio of $1.17\pm 0.10$, where the estimate
of the uncertainty takes into account that the window is
Gaussian.  Assuming a CR flux of $\Phi_{CR}(E) = 30 (E/{\rm
EeV})^{-3} {\rm ~EeV^{-1}\ km^{-2}\ yr^{-1}\ sr^{-1}}$ we can
then calculate the 95\,\% confidence limits (CL) for the upper
bound on the number of events from the source to be
$\Phi_{s}^{95}(E > 10^{17.9} {\rm ~eV}) = 0.04 \ {\rm km^{-2}\
yr^{-1}}$.  This upper limit is more than an order of magnitude
below predictions made for neutron fluxes from the GC
\cite{bo03,ah05} and is at the level of the prediction made in
Ref.\ \cite{gr05}.

\section{Upper limit of the photon fraction}

As mentioned above, photon primaries are expected to dominate
over nucleon primaries in non-acceleration (``top-down'') models
of EHECR origin.\cite{Bhattacharjee00}
Thus, the determination of the photon contribution is a crucial
probe of cosmic-ray source models.  Separating photon-induced
showers from events initiated by nuclear primaries is
experimentally much easier than distinguishing light and heavy
nuclear primaries.  As an example, average depths of shower
maxima at 10~EeV primary energy are predicted to be about
1000~g~cm$^{-2}$, 800~g~cm$^{-2}$, and 700~g~cm$^{-2}$ for
primary photons, protons, and iron nuclei, respectively.
Moreover, analyses of nuclear composition are uncertain due to
our poor knowledge of hadronic interactions at very high
energies.  Photon showers, being driven mostly by electromagnetic
interactions, are less affected by such uncertainties and can be
modelled with greater confidence.

So far limits on the UHE photon fraction in cosmic rays have been
set by ground arrays only.  By comparing the rates of
near-vertical showers to inclined ones recorded by the Haverah
Park shower detector, upper limits (95\% CL) of 48\% above
10~EeV and 50\% above 40~EeV were deduced.\cite{ave} Based on an
analysis of muons in air showers observed by AGASA, the upper
limits (95\% CL) to the photon fraction were estimated to be 28\%
above 10~EeV and 67\% above 32~EeV.\cite{shinozaki} An upper
limit of 67\% (95\% CL) above 125~EeV was derived in a
dedicated study of the highest energy AGASA
events.\cite{risse05}

The fluorescence telescopes of the Pierre Auger Observatory are
ideal instruments for such an analysis, since they measure the
position of the shower maximum $X_{\rm max}$ as the
discriminating observable directly.  To achieve a high accuracy
in reconstructing the shower geometry, we make use of the
``hybrid'' detection technique, i.e.\ we select events observed
by both the ground array and the fluorescence
telescopes.\cite{Kampert06}

Compared to air showers initiated by nuclear primaries, photon
showers at energies above 10~EeV are in general expected to have
a larger depth of shower maximum $X_{\rm max}$ and to contain
fewer secondary muons.  The latter is because the mean free paths
for photo-nuclear interactions and direct muon pair production
are more than two orders of magnitude larger than the radiation
length.  Consequently, only a small fraction of the primary
energy in photon showers is generally transferred into secondary
hadrons and muons.  The large $X_{\rm max}$ values for photon
showers at 10~EeV are essentially due to the small multiplicity
in electromagnetic interactions, in contrast to the large number
of secondaries produced in inelastic interactions of high-energy
hadrons.  Secondly, because of the LPM effect\cite{lpm}, the
development of photon showers is even further delayed above
$\sim$~10~EeV. Another feature of the LPM effect is an increase
in shower fluctuations: $X_{\rm max}$ fluctuations for photon
showers are $\sim$~80~g~cm$^{-2}$ at 10~EeV, compared to
$\sim$~60~g~cm$^{-2}$ and $\sim$~20~g~cm$^{-2}$ for primary
protons and iron nuclei, respectively.  At higher energies,
cosmic-ray photons may convert in the geomagnetic field and
create a pre-shower before entering the atmosphere.  The energy
threshold for geomagnetic conversion is $\sim$~50~EeV for the
Auger southern site.  Conversion probability and pre-shower
features depend both on primary energy and arrival direction.  In
the case of a pre-shower, the subsequent air shower is initiated
as a superposition of lower-energy secondary photons and
electrons.  For air showers from converted photons, the $X_{\rm
max}$ values and the fluctuations are considerably smaller than
from single photons of same total energy.  From the point of view
of air shower development, the LPM effect and pre-shower
formation compete with each other.  The cascading of photons in
the geomagnetic field is simulated with the PRESHOWER
code~\cite{homola} and the shower development in air, including
the LPM effect~\cite{lpm}, is calculated with
CORSIKA~\cite{heck}.  For photo-nuclear processes, an
extrapolation of the cross-section as given by the Particle Data
Group has been employed.\cite{pdg} QGSJET~01 has been used as a
hadron event generator.~\cite{qgs01}

The Auger data used in this analysis were taken with a total of
12 fluorescence telescopes situated a two sites and with the
number of surface detector stations growing during this period
from about 150 to 950.  For the present analysis, we selected
hybrid events, i.e.\ showers observed both with (one or more)
surface tanks and telescopes.  Even when only one tank is
triggered, the reconstruction of the shower geometry and thereby
of $X_{\rm max}$ improves strongly.\cite{Kampert06} The
reconstruction of the shower profiles accounts for the time
varying atmospheric density profiles, aerosol concentrations, and
cloud coverage.  After subtracting the Cherenkov light
contribution, a Gaisser-Hillas function \cite{gh} is fitted to
the profile to obtain the depth of shower maximum and the
calorimetric shower energy is obtained by integration.  The
quality cuts applied for event selection and further details of
this analysis are given in Ref.\cite{photonlimit}.

\begin{figure}[t]
    \centerline{\epsfxsize=9cm\epsfbox{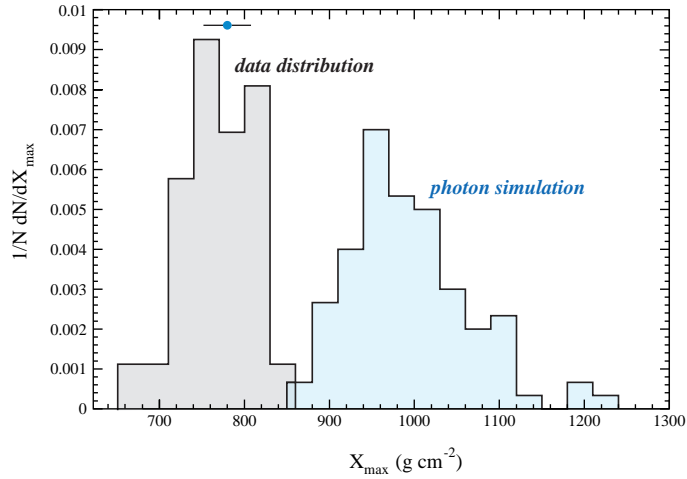}}
    \vspace*{-3mm}\caption[xx]{$X_{\rm max}$ distribution of
    experimental and simulated photon events.  The point with
    error bar represents the $X_{\rm max}$ value and its
    uncertainty of one data event.  For each measured event, 100
    photon induced showers were generated taking into account the
    arrival direction and energy of the data event.  The result 
    of the simulations for the event shown is represented by
    the blue histogram.
    \label{fig:xmaxcomparison}}
\end{figure}
 
\begin{figure}[t]
    \centerline{\epsfxsize=11cm\epsfbox{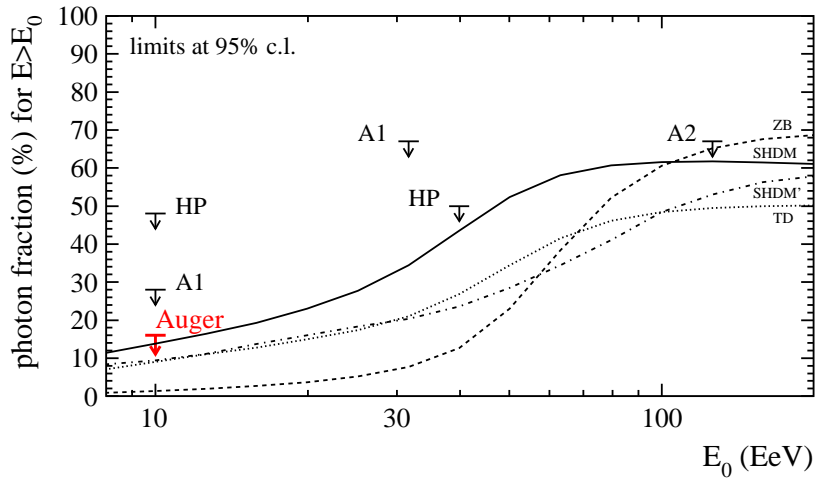}}
    \vspace*{-4mm}\caption[xx]{Upper limits (95\% CL) to the
    cosmic-ray photon fraction derived from the Pierre Auger
    experiment and obtained previously from AGASA
    (A1)~\cite{shinozaki}, (A2)~\cite{risse05} and Haverah Park
    (HP)~\cite{ave} data, compared to expectations for
    non-acceleration models (ZB, SHDM, TD from~\cite{models},
    SHDM' from \cite{ellis}).\cite{photonlimit}
    \label{fig:photonlimits}}
\end{figure}

After applying the strong selection cuts to the data, 29 events
with energies above 10~EeV remained for the analysis.  The
$X_{\rm max}$ distribution of these events is displayed in
Figure~\ref{fig:xmaxcomparison}.  The single point with error bar
represents the $X_{\rm max}$ value and its uncertainty of one
typical data event.  For each of such events, 100 photon induced
showers were generated taking into account the arrival direction
and energy of that data event.  The resulting photon expectation
for that single event is represented by the blue histogram.  The
present $X_{\rm max}$ uncertainties are conservative estimates
and are expected to decrease significantly in the future.  The
main contributions are the profile fit, the atmospheric
conditions, and the shower geometry.\cite{photonlimit} For all
events, the observed $X_{\rm max}$ is well below the average
value expected for photons.  Differences between photon
predictions and data range from +2.0 to +3.8 standard deviations.
Taking the available statistics, the individual differences
between data and photon predictions, and the systematic
uncertainties of data and simulations into account, an upper
limit of the photon fraction of 16\,\% at 95\,\% CL is derived.
This is plotted in Figure \ref{fig:photonlimits} together with
previous experimental limits and some illustrative estimates for
non-acceleration models.  The derived limit is the first one
based on observing the depth of shower maximum with the
fluorescence technique.  The result confirms and improves
previous limits above 10~EeV that came from surface arrays.  It
is worth mentioning that this improved limit is achieved with
only 29 events above 10~EeV, as compared to about 50 events in
the Haverah Park analysis and about 120 events in the AGASA
analysis.  In the very near future and with increasing
statistics, the limit can be reduced by at least a factor of
three at 10~EeV and limits will be set also at higher energies
constraining models significantly.

\section{First estimate of the energy spectrum}

A major goal of the Pierre Auger Observatory is to make a
reliable measurement of the cosmic-ray energy spectrum above 10
EeV and to answer the question about the existence of the GZK
cut-off.  The large aperture of the Auger surface array will
allow for the first time an observation of the CR spectrum in
this energy range with good statistics.  Moreover, the hybrid
design will allow to resolve the discrepancy between previous
spectrum measurements that were based on the different
techniques.

The Pierre Auger measurement profits from the hybrid technique of
the ground array and fluorescence telescopes.  The 100\,\% duty
cycle of the ground array provides sufficient statistics, even
though the analyzed data presented here (taken from 01/01/2004
through 06/05/2005) correspond to less than 4 months of the
amount we anticipate, once the Auger South array will be
completed.  The exposure available for this analysis is 1750
km$^{2}$ steradian years, slightly larger than that achieved by
AGASA. The fluorescence information available for a subset of
showers observed in hybrid mode allows to determine the absolute
energy scale.  The energy estimate of the ground array uses the
signal size at a radius of 1000 m from the shower core (
``$S(1000)$'' ), which is determined from a fit to the lateral
distribution of signal sizes from all the tanks triggered by an
air shower.  The ``Constant Intensity Cut'' method~\cite{so05} is
used to re-scale values from different shower inclinations.
$S(1000)$ is almost linearly proportional to the energy of the
primary particle.  The conversion factor that relates $S(1000)$
to the energy is experimentally determined from the hybrid events
by use of the very good energy reconstruction based on the
fluorescence detector information.  This reduces significantly
the dependence on air shower models \cite{Drescher04} and on
assumptions of the UHECR composition, compared to previous
surface array experiments.

\begin{figure}[t]
\begin{minipage}[b]{0.48\textwidth}
\mbox{}\\
\centerline{\epsfxsize=6.1cm\epsfbox{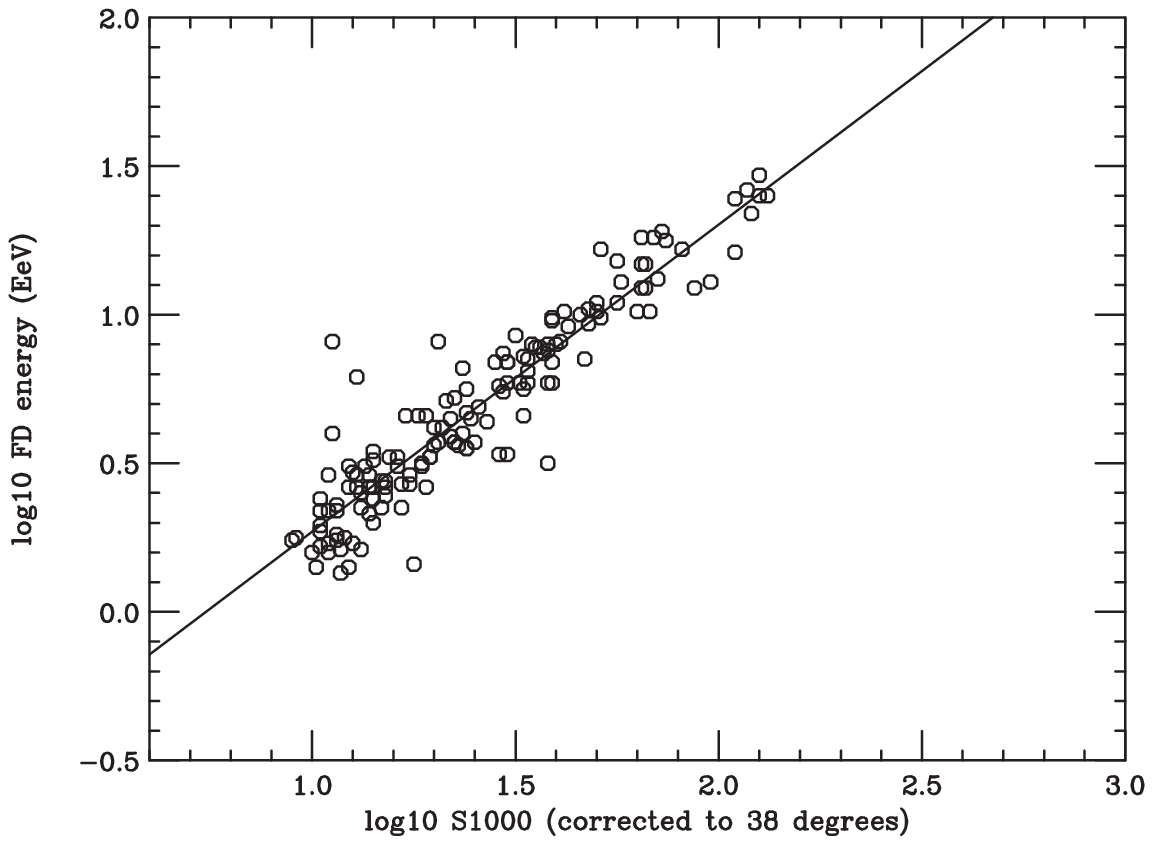}}
\caption[xx] {Energy measured by the flourescence detector vs 
$S(1000)$ as measured by the ground array for hybrid events with 
zenith angles $<$~$60^{\circ}$.}
\label{fig:fd-vs-sd}    
\end{minipage}
\hfill
\begin{minipage}[b]{0.47\textwidth}
\mbox{}\\
\centerline{\epsfxsize=7.1cm\epsfbox{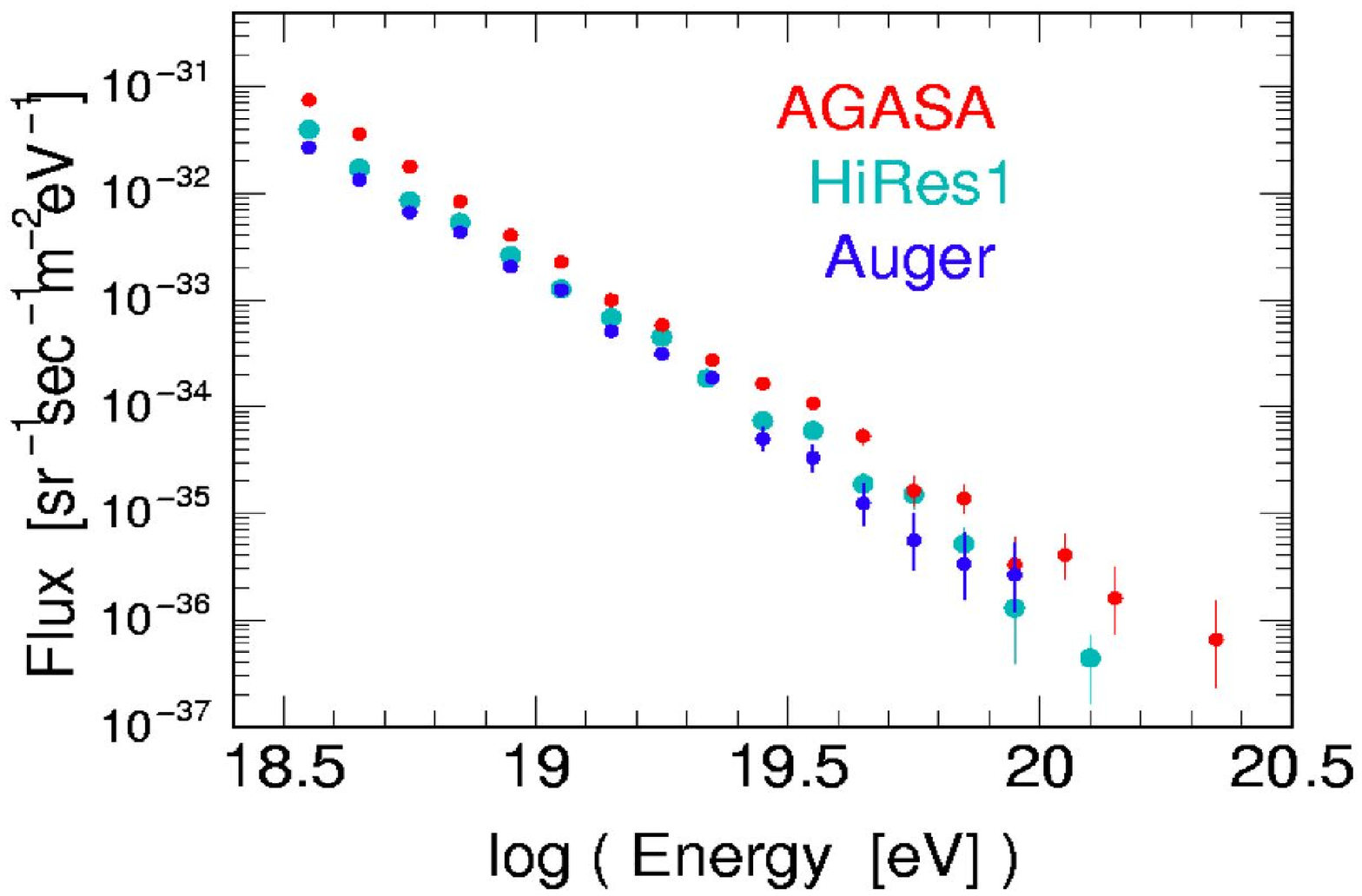}}
\caption[energy spectra Auger, AGASA, HiRes-I]{ The energy
spectrum of EHECRs measured by the Pierre Auger experiment
compared with the AGASA \cite{agasa_spec} and
HiRes-I~\cite{hires_spec} results.}
\label{fig:augerspeccomp}
\end{minipage}
\end{figure}

Figure~\ref{fig:fd-vs-sd} shows the correlation between $S(1000)$
and the energy determined from the fluorescence telescopes.  Even
though the data are still at a very preliminary stage and the
reconstruction procedures are still to be improved, the
correlation is very convincing.  Figure~\ref{fig:augerspeccomp}
compares the obtained energy spectrum with those from AGASA and
HiRes-I.\cite{agasa_spec,hires_spec} ~Our data points contain
around 3500 events above 3 EeV. Above this energy, the full
geometrical area of the detector, defined by the layout of the
water tanks, is sensitive so that determination of the flux of
events is relatively straightforward.  The general form is
similar to the earlier experiments but, even allowing for the
systematic uncertainties still present, it appears that at the
highest energies significantly fewer events are seen than
expected from the AGASA analysis.  The claim of the HiRes team
that the spectrum steepens at the highest energies can neither be
confirmed nor denied with the present exposure.  One event was
recorded in April 2004 for which the fluorescence reconstruction
gives an energy greater than 140 EeV, but the particle array was
small at that date and the shower core fell outside of the
fiducial area.  Details of the spectrum will be greatly clarified
with the data that have been accumulated since June 2005.

\section{Summary and Outlook}

The construction of the southern Pierre Auger Observatory is well
underway.  About 1000 stations of the surface array and 18
telescopes of the fluorescence detector are in operation and
taking data routinely.  Completion of the southern site is
planned for 2007 and R\&D work for the northern site to be
located in south east Colorado has started.

Parallel to the completion of the observatory, first science
results were already obtained on the energy spectrum, searches
for localized anisotropies near the direction of the Galactic
Center, and on setting upper limits on the photon fraction of the
primary particles.  Considering the very limited statistics from
1.5 years of data taking during construction, being equivalent of
only 3 months of a full array, it is not surprising that the
emergence of a clear picture about the shape of the energy
spectrum above the GZK threshold needs a little more time.  The
2.3 years of data used for the anisotropy searches in the
Galactic Center region provides statistics much greater than
those of previous experiments.  No evidence for a point-like
source in the direction of Sagittarius\,A$^{*}$ was found.  This
excludes several scenarios of neutron sources in the GC suggested
recently.  Our searches on larger angular windows in the
neighborhood of the GC do not show abnormally over-dense regions.
In particular, they do not support the large excesses reported in
AGASA data (of 22\% on $20^\circ$ scales) and SUGAR data (of 85\%
on $5.5^\circ$ scales).  The upper limit to the photon fraction
above 10~EeV derived from a direct observation of the shower
maximum confirms and reduces previous limits from ground arrays.
Again, the current analysis is limited mainly by the small number
of events.

\begin{figure}[t]
    \centerline{\epsfxsize=8cm\epsfbox{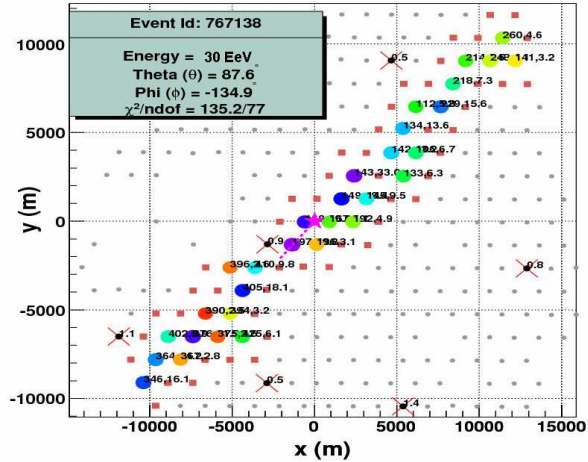}}
    \vspace*{-3mm}\caption[xx]{Example of a near horizontal 
    air shower as seen by the ground array. The shower has 
    triggered 31 stations and extends over 30 km at 
    ground.
    \label{fig:inclined}}
\end{figure}

The number of hybrid events will considerably increase over the
next years, allowing to set much stronger limits on the
anisotropy and point source searches and on the photon limits.
It will also reduce the uncertainties of the energy spectrum and
will allow for further studies of EHECRs.  For example, the
Pierre Auger Collaboration is developing the study of inclined
events, and showers with zenith angles above $85^{\circ}$ have
been seen.  This was expected as they had been detected long ago
with much smaller arrays, but the richness of the new data is
impressive.  Figure \ref{fig:inclined} shows an event at about
$88^{\circ}$ with 31 detectors, and even the present array is too
small to contain it.  A preliminary estimate of its energy is
around 30 EeV. An understanding of these events will lead to
additional aperture for collection of the highest-energy
particles and also give additional routes to understanding the
mass composition.  Further, these events form the background
against which a neutrino flux might be detectable.  There is an
exciting future ahead.

\vspace*{5mm} {\small {\bf Acknowledgement:} It is a pleasure to
thank the organizers of the Symposium for their invitation to
this Symposium.  The work of the group at University Wuppertal is
supported in part by the German Ministry for Research and
Education (Grant 05 CU5PX1/6).}

\end{document}